\documentclass[aps,prl,twocolumn,superscriptaddress,floatfix]{revtex4}%
\usepackage{graphicx}
\usepackage{subfigure}
\usepackage{dcolumn}
\usepackage{bm}
\usepackage{hyperref}
\usepackage{amssymb}
\usepackage{amsmath}
\usepackage{amsfonts}
\usepackage{amssymb}
\usepackage{color}%
\begin{document}
\title{Broad Feshbach resonance with a large background scattering length in a fermionic atom-molecule mixture}
\author{Zhen Su}
\thanks{These authors contributed equally to this work.}
\affiliation{Hefei National Research Center for Physical Sciences at the Microscale and School of Physical Sciences, University of Science and Technology of China, Hefei 230026, China}
\affiliation{Shanghai Research Center for Quantum Science and CAS Center for Excellence in Quantum Information and Quantum Physics, University of Science and Technology of China, Shanghai 201315, China}
\author{Tong-Hui Shou}
\thanks{These authors contributed equally to this work.}
\affiliation{Hefei National Research Center for Physical Sciences at the Microscale and School of Physical Sciences, University of Science and Technology of China, Hefei 230026, China}
\affiliation{Shanghai Research Center for Quantum Science and CAS Center for Excellence in Quantum Information and Quantum Physics, University of Science and Technology of China, Shanghai 201315, China}
\author{Huan Yang}
\thanks{These authors contributed equally to this work.}
\affiliation{Hefei National Research Center for Physical Sciences at the Microscale and School of Physical Sciences, University of Science and Technology of China, Hefei 230026, China}
\affiliation{Shanghai Research Center for Quantum Science and CAS Center for Excellence in Quantum Information and Quantum Physics, University of Science and Technology of China, Shanghai 201315, China}
\affiliation{Hefei National Laboratory, University of Science and Technology of China, Hefei 230088, China}
\author{Jin Cao}
\affiliation{Hefei National Research Center for Physical Sciences at the Microscale and School of Physical Sciences, University of Science and Technology of China, Hefei 230026, China}
\affiliation{Shanghai Research Center for Quantum Science and CAS Center for Excellence in Quantum Information and Quantum Physics, University of Science and Technology of China, Shanghai 201315, China}

\author{Bo-Yuan Wang}
\affiliation{Hefei National Research Center for Physical Sciences at the Microscale and School of Physical Sciences, University of Science and Technology of China, Hefei 230026, China}
\affiliation{Shanghai Research Center for Quantum Science and CAS Center for Excellence in Quantum Information and Quantum Physics, University of Science and Technology of China, Shanghai 201315, China}

\author{Ting Xie}
\affiliation{Hefei National Research Center for Physical Sciences at the Microscale and School of Physical Sciences, University of Science and Technology of China, Hefei 230026, China}
\affiliation{Shanghai Research Center for Quantum Science and CAS Center for Excellence in Quantum Information and Quantum Physics, University of Science and Technology of China, Shanghai 201315, China}
\affiliation{Hefei National Laboratory, University of Science and Technology of China, Hefei 230088, China}
\author{Jun Rui}
\affiliation{Hefei National Research Center for Physical Sciences at the Microscale and School of Physical Sciences, University of Science and Technology of China, Hefei 230026, China}
\affiliation{Shanghai Research Center for Quantum Science and CAS Center for Excellence in Quantum Information and Quantum Physics, University of Science and Technology of China, Shanghai 201315, China}
\affiliation{Hefei National Laboratory, University of Science and Technology of China, Hefei 230088, China}
\author{Bo Zhao}
\affiliation{Hefei National Research Center for Physical Sciences at the Microscale and School of Physical Sciences, University of Science and Technology of China, Hefei 230026, China}
\affiliation{Shanghai Research Center for Quantum Science and CAS Center for Excellence in Quantum Information and Quantum Physics, University of Science and Technology of China, Shanghai 201315, China}
\affiliation{Hefei National Laboratory, University of Science and Technology of China, Hefei 230088, China}
\author{Jian-Wei Pan}
\affiliation{Hefei National Research Center for Physical Sciences at the Microscale and School of Physical Sciences, University of Science and Technology of China, Hefei 230026, China}
\affiliation{Shanghai Research Center for Quantum Science and CAS Center for Excellence in Quantum Information and Quantum Physics, University of Science and Technology of China, Shanghai 201315, China}
\affiliation{Hefei National Laboratory, University of Science and Technology of China, Hefei 230088, China}

\begin{abstract}{We report the observation of a broad magnetic Feshbach resonance with a large background scattering length in an ultracold fermionic mixture of $^{23}$Na$^{40}$K molecules and $^{40}$K atoms, with both species prepared in their lowest hyperfine states. The Feshbach resonance is characterized by measuring resonantly enhanced loss rates and elastic scattering cross sections via cross-species thermalization. The large background scattering length can drive the atom-molecule mixture into the hydrodynamic regime when the magnetic field is far from the resonance. We observe that the center-of-mass motions of the atoms and molecules are phase-locked and oscillate with a common frequency due to hydrodynamic drag effects. This broad atom-molecule Feshbach resonance with its large background scattering length opens up a new avenue towards studying strongly interacting fermionic gases with mass imbalance.}

\end{abstract}
\maketitle

Ultracold Fermi gases with tunable interactions have emerged as a powerful platform for exploring quantum many-body physics. In recent years, remarkable advances have been achieved in the study of BCS-BEC crossover and strongly interacting fermionic superfluids \cite{Bloch2008,Giorgini2008,chin2010,Chenq2024}. These research efforts have predominantly utilized two internal states of a single fermionic species, with interactions controlled via s-wave magnetic Feshbach resonances. Building upon these achievements, the focus has now spread to fermionic mixtures, which introduce mass imbalance as an additional degree of freedom \cite{Baarsma2010,Kohstall2012,Cosetta2024}. This configuration offers a unique opportunity to investigate novel pairing mechanisms, such as the Fulde-Ferrell-Larkin-Ovchinnikov state and breached pairing superfluid \cite{Gubankova2003,Forbes2005,Fulde1964,Larkin1965,Radzihovsky2010}, thereby expanding the frontiers of fermionic superfluids.

Investigating strongly interacting fermionic gases critically relies on the broad Feshbach resonances with extensive universal ranges. The interplay between a broad resonance and the Pauli exclusion principle ensures collisional stability near the resonance, a feature that is essential to investigate strongly interacting gases \cite{Petrov2004, Petrov2005}.  Both  $^6$Li and $^{40}$K atoms possess such broad Feshbach resonances. Particularly, the large background scattering length relevant to the broad resonance in $^6$Li makes it the most favorite option for studying
fermionic superfluids \cite{Abraham1997, Ketterle2008,chin2010}. However, broad resonances are rare in mass-imbalanced mixtures of fermionic atoms. Recently, experimental surveys of Feshbach resonances between various fermionic atomic species have been conducted, encompassing mixtures such as $^{6}$Li and $^{40}$K \cite{Wille2008, Tiecke2010}, $^{161}$Dy and $^{40}$K \cite{Ravensbergen2020}, $^{6}$Li and $^{173}$Yb \cite{Green2020}, $^{6}$Li and $^{53}$Cr \cite{Finelli2024}, and $^{6}$Li and $^{167}$Er \cite{Florian2023}. Although these investigations have identified numerous resonances, the majority of them are narrow resonances. Only a broad Feshbach resonance is identified in the $^{161}$Dy and $^{40}$K mixture \cite{Ravensbergen2020}, which offers the potential to explore strongly interacting fermionic gases.

The advent of ultracold molecules provides new opportunities to study quantum many-body physics \cite{Ni2008,Takekoshi2014,Molony2014,Park2015,Guo2016,Voges2020,Stevenson2023,He2024}. While most research has focused on dipolar interactions in pure molecular gases \cite{Giacomo2020,Matsuda2020,Schindewolf2022,Lysander2023,Bigagli2024}, atom-molecule mixtures have attracted increasing attention for investigating quantum mixtures \cite{Yang2019,Wang2021, Son2022,Nichols2022,Park2023b}. Especially, the observation of magnetic Feshbach resonances between $^{40}$K atoms and $^{23}$Na$^{40}$K molecules have suggested this fermionic mixture as a possible candidate for studying mass-imbalanced strongly interacting fermionic gases \cite{Yang2019,Wang2021}.
However, a key challenge persists: most observed resonances involve collision channels where either atoms or molecules are not in their lowest hyperfine state. To date, only a few narrow resonances have been identified in the lowest hyperfine combination. Thus, the existence of a viable resonance for investigating strongly interacting fermionic mixture remains elusive.

In this Letter, we report the observation of a broad Feshbach resonance at 217.3 G with a large background scattering length between ultracold $^{23}$Na$^{40}$K molecules and $^{40}$K atoms in their lowest hyperfine states.  The resonance is characterized by measuring the loss rates and elastic scattering cross sections. We find that this resonance is highly open-channel dominated due to the large background scattering length of $a_{\rm{bg}}=-2902$ $a_0$ and a width of $\Delta=23.1$ G. The resonance strength, $A=a_{\rm bg}\Delta/a_0\approx 66000$ G, is larger than that of the broad Feshbach resonance between $^{161}$Dy and $^{40}$K by about one order of magnitude \cite{Ravensbergen2020}. The background scattering length is so large that when the magnetic field is far from the resonance, the system can still be driven into the hydrodynamic regime. We have observed that the center of mass dynamics of the molecules and atoms along the gravitational direction, associated with the highest trap frequency, are phase-locked due to the strong atom-molecule interactions. This broad Feshbach resonance provides a promising avenue for exploring strongly interacting mass-imbalanced fermionic mixtures.

Our experiment starts with the preparation of an ultracold mixture of $^{23}\text{Na}^{40}\text{K}$ molecules and $^{40}\text{K}$ atoms confined in an optical dipole trap. The experimental setup and the preparation procedures have been described minutely in previous works \cite{Yang2022,Su2022}. In brief, we first prepare a quantum degenerate mixture of about $3\times 10^5$ $^{23}\text{Na}$ and  about $2.8\times 10^5$ $^{40}\text{K}$ atoms at approximately 300 nK.  We create about $1.4\times 10^4$ $^{23}\text{Na}^{40}\text{K}$ molecules through magnetoassociation and stimulated Raman adiabatic passage at 77.6 G. The $^{23}\text{Na}^{40}\text{K}$ molecules are prepared in the absolute ground state $|v,N,m_{\mathrm{Na}},m_{\mathrm{K}}\rangle=|0,0,\frac{3}{2},-4\rangle$, where $v$ represents the vibrational quantum number, $N$ is the rotational quantum number, and $m_{\mathrm{Na}}$ and $m_{\mathrm{K}}$ denote the nuclear spin projections of $^{23}\text{Na}$ and $^{40}\text{K}$, respectively. The $^{40}\text{K}$ atoms are prepared in the lowest hyperfine state $|F,m_{\mathrm{F}}\rangle=|\frac{9}{2},-\frac{9}{2}\rangle$.
%In the atom-molecule mixture, the number of $^{40}\text{K}$ atoms is approximately an order of magnitude larger than that of $^{23}\text{Na}^{40}\text{K}$ molecules.
The trap frequencies for $^{40}\text{K}$ atoms are $2\pi\times(71,265,24)$ Hz, where the strongest confinement is along the gravitational direction. In our experiment, the typical molecular lifetime in the atom-molecule mixture is about 40 ms, and that of the pure molecular gas is about one order of magnitude larger.

We first locate the resonance position by ramping the magnetic field to a target value and observe enhanced loss features \cite{supp}.
%within 15 ms. The mixture is held at the magnetic field for 20 ms. The $^{40}$K atoms are then removed using a resonant laser pulse, and the magnetic field is ramped back to 77.6 G in 5 ms. The remaining $^{23}\text{Na}^{40}\text{K}$ molecules are then transferred back to the Feshbach state and detected via absorption imaging.
The residual number of molecules after a hold time of 5 ms at different magnetic fields are shown in Fig. 1. A broad Feshbach resonance located at about 217 G is observed, together with two well-separated narrow resonances at about 191 and 232 G identified. We measure the loss rate coefficients close to the broad resonance position by monitoring the time evolution of the number of molecules at the target magnetic field.  Although two-body inelastic collisions are energetically forbidden for particles in their absolute ground states, the trapping laser can excite the triatomic collision complex thereby causing two-body collisional loss \cite{Christianen2019,Nichols2022,Yang2022b,Cao2024}. We determine the loss rate coefficients by the ratio of the decay rate to the density of the atom cloud overlapped with the molecule cloud. The measured loss rate coefficients versus magnetic fields are shown in the inset of Fig. 1. A Lorentzian fit gives the resonance position $B_0=217.3(4)$ G. The measured peak loss rate coefficient is about 3 times larger than the universal loss rate coefficient of $1.3\times10^{-10}$cm$^{3}$/s.

\begin{figure}[ptb]
\centering
\includegraphics[width=8cm]{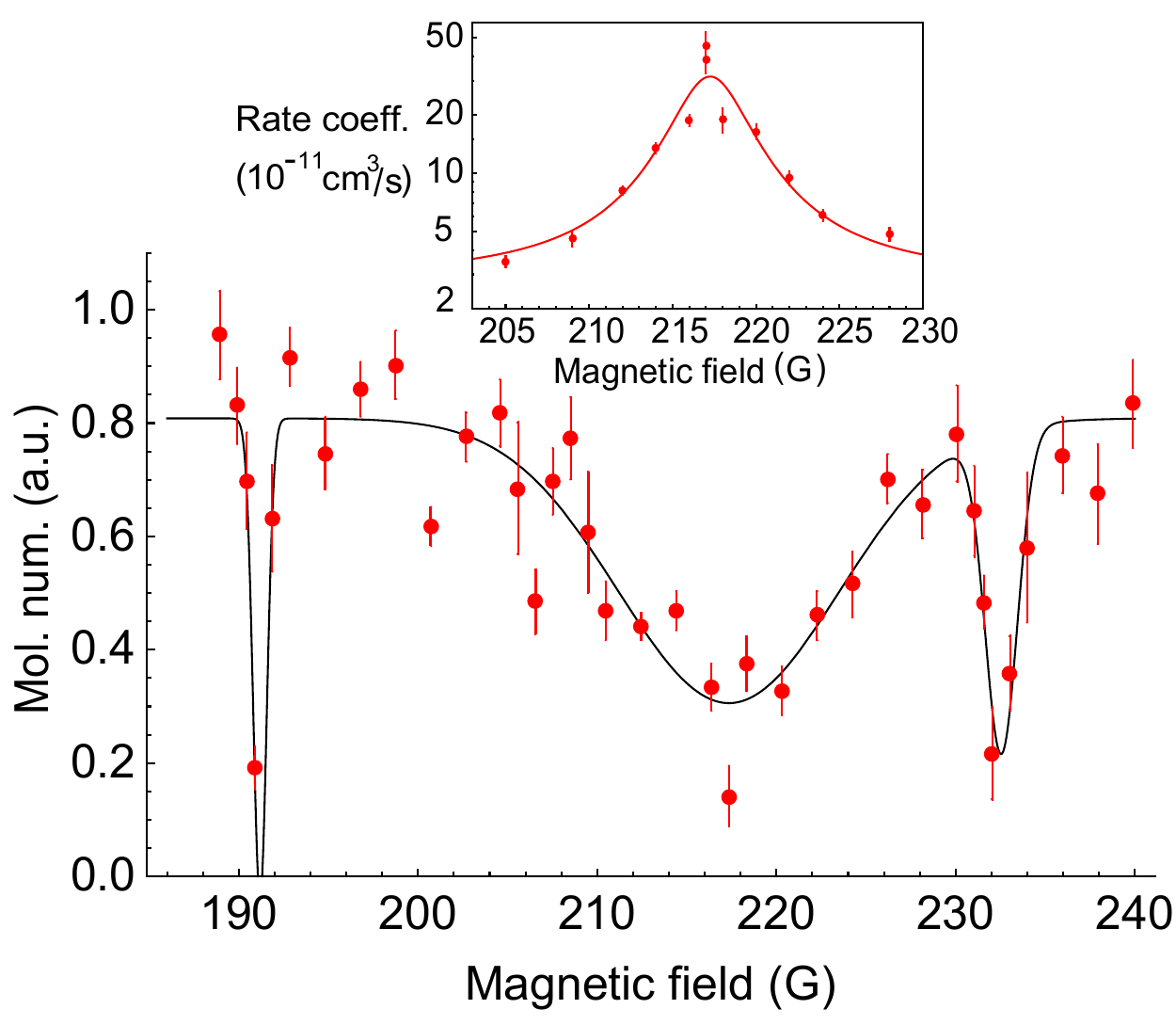}\caption{The location of the broad atom-molecule Feshbach resonance. The number of remaining $^{23}$Na$^{40}$K molecules is shown as a function of the magnetic field. The black solid line is a Gaussian fit to the data. The inset shows the loss rate coefficients in the vicinity of the broad Feshbach resonance. The red solid line is a Lorentzian fit to the data. The fit yields the resonance position $B_0=217.3(4)$ G. Error bars shown in the plot are the standard errors.}
\label{Fig1.lossratecoeff.pdf}%
\end{figure}

To further characterize this broad Feshbach resonance, we measure the elastic scattering cross sections between atoms and molecules via cross-species thermalization \cite{Mosk2001,Son2020,Tobias2020,Su2022}.  In the experiment, after ramping the magnetic field to a target value, we apply a 30 $\mu$s weak and resonant light pulse to heat the $^{40}$K atoms to about 450 nK.  The heating pulse also reduces the number of $^{40}$K atoms to about $1.0\times 10^5$. The mixture is then held at the target field for variable durations. Due to the elastic collisions between $^{23}$Na$^{40}$K and $^{40}$K, the temperature of the molecules will increase during this hold time. Since the number of atoms is much larger than that of the molecules, we assume that the temperature of $^{40}$K remains a constant. The variation of the temperature of $^{23}$Na$^{40}$K can be described by the equation $T(t)=T_f-(T_f-T_0)e^{-\Gamma_{\rm{th}}t}$, where $T_0$ and $T_f$ represent the respective initial and final temperatures and $\Gamma_{\rm{th}}$ is the thermalization rate. %{\color{red}The cloud sizes along the $x$ and $y$ directions after a time of flight can be described by  %$\sigma_{x,y}^2(t)=[k_{\rm{B}}T(t)/m_{\rm{NaK}}](\tau_{\rm{tof}}^2+1/\omega_{x,y}^2)$, with $\tau_{\rm{tof}}$ being the time of flight.}
Assuming the density distribution of the molecular cloud is described by  a Gaussian function, the square of the molecular cloud size after a time of flight $t_{\rm{tof}}$ may be described by $\sigma_{x,y}^2(t)=[k_{\rm{B}}T(t)/m_{\rm{NaK}}](t_{\rm{tof}}^2+1/\omega_{x,y}^2)$, allowing us to determine the thermalization rate by monitoring the size of the molecular cloud. We measure the molecular cloud size after a time of flight of 3-4 ms. As shown in Fig. 2, the thermalization can be recognized from the molecular cloud size as a function of the hold time. Remarkably the thermalization rate is reduced evidently near 194 G, which suggests it is close to the minimum of the scattering cross sections. Such a minimum is caused by the Fano interference between the continuum states in the open channel and the discrete state in the closed channel. Close to the resonance, the loss is strong, resulting in a molecular lifetime of merely 3-4 ms. This is much shorter than the typical hold time of several tens of milliseconds necessary for thermalization measurements and thus the thermalization process cannot be measured. To extract the thermalization rate, we employ the function $\sigma_{x,y}^2(t)=A_{x,y}e^{-\Gamma_{x,y} t}+B_{x,y}$ to fit the data points, with $x$ and $y$ denoting the horizontal and vertical directions, respectively. The thermalization rate is then obtained by $\Gamma_{\rm{th}}=(\Gamma_x+\Gamma_y)/2$.

\begin{figure}[ptb]
\centering
\includegraphics[width=8cm]{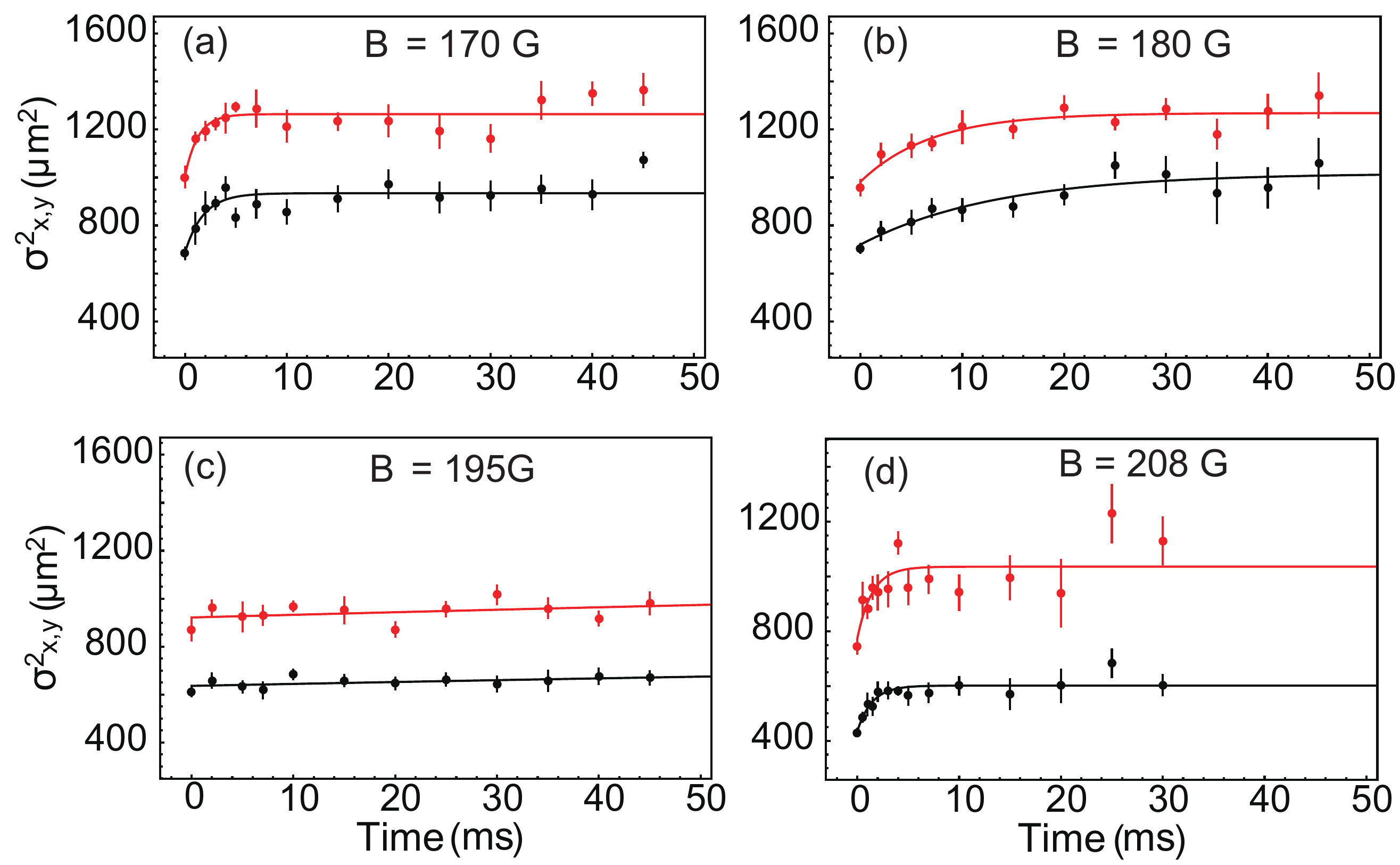}\caption{Cross-species thermalization measurements at different magnetic fields. The square of the size of the molecular cloud evolves as a function of hold time after heating the atoms by a resonant heating pulse. The red and black points represent the data of $\sigma_x^2$ and $\sigma_y^2$ respectively, and the solid lines represent the fits of the data to the model described in the main text. Error bars shown in the plots are standard errors of the mean.}
\label{Fig2.thermaldata.pdf}
\end{figure}

The elastic scattering cross sections are extracted from the thermalization measurements using the formula $\Gamma_{\rm{th}}= \Gamma_{\rm{coll}}/(3/\xi)$, where $\Gamma_{\rm{coll}}=n_{\rm{ov}}\sigma_{\rm{el}} v_{\rm{rel}}$ is the elastic collision rate \cite{Mosk2001,Son2020,Tobias2020,Su2022}. Here, $n_{\rm{ov}}=(N_{\rm{NaK}}+N_{\rm{K}})[(\frac{2\pi k_{\rm{B}} T_{\rm{K}}}{m_{\rm{K}} \bar\omega_{\rm{K}}^2})(1+\frac{m_{\rm{K}}T_{\rm{NaK}}}{m_{\rm{NaK}} T_{\rm{K}} \gamma_t^2} )]^{-\frac{3}{2}}$ represents the overlap density with $\gamma_t\approx0.77$ being the ratio between the trap frequencies for $^{23}$Na$^{40}$K molecules and $^{40}$K atoms, $v_{\rm{rel}}=\sqrt{(8 k_{\rm{B}}/\pi)(T_{\rm{K}}/m_{\rm{K}}+T_{\rm{NaK}}/m_{\rm{NaK}})}$ is the relative mean velocity, and the parameter $\xi=4m_{\rm{NaK}}m_{\rm{K}}/(m_{\rm{NaK}}+m_{\rm{K}})^2$ accounts for the mass effect. The elastic scattering cross sections determined in this way are shown in Fig. 3 as a function of magnetic field.

We obtain the properties of the broad resonance around 217 G by the following procedures. Assuming the momentum-dependent factor can be neglected, the elastic scattering cross sections is proportional to the absolute value of the square of scattering length. Therefore, we have $\sigma=4\pi|a|^2$, and a simple formula can describe the \emph{s}-wave scattering length \cite{supp}
\begin{equation}
\begin{aligned}
a(B)=a_{\rm{bg}}\left(1-\frac{\Delta}{B-B_{0}-i \gamma/2}\right),
%|\beta(B)|^2=\frac{a_{bg}^2\,\Delta^2}{(B-B_0)^2+\frac{\gamma^2}{4}}
\label{eq1}
\end{aligned}
\end{equation}
where $a_{\mathrm{bg}}$ is the background scattering length, $\Delta$ denotes the resonance width, $B_0$ is the location of the resonance, and $\gamma$ represents the decay of the bound state \cite{Hutson2007}. Close to the minimum of scattering length, $a(B)$ is significantly reduced, allowing us to safely omit the momentum-dependent factor. Therefore, we fit the data points between 180 G and 202 G to $\sigma=4\pi|a|^2$, where the measured scattering cross sections are smaller than $10^{-9}$ cm$^{2}$. The resonance position is fixed at $B_0=217.3$ G, and $a_{\mathrm{bg}}$, $\Delta$ and $\gamma$ are fitting parameters. We obtain $\Delta=-23.1(2)$ G, $a_{\mathrm{bg}}=-2902(182) a_0$, and $\gamma=2.7(5)$ G. Note that the background scattering length is fundamental nature of the potential energy curve of the collision channel. The fit of the elastic scattering cross sections can only determine its absolute value. Its sign is determined by the magnetic field where the scattering cross section reaches its minimum and by the direction from which the bound state approaches the resonance \cite{supp}.

\begin{figure}[ptb]
\centering
\includegraphics[width=8cm]{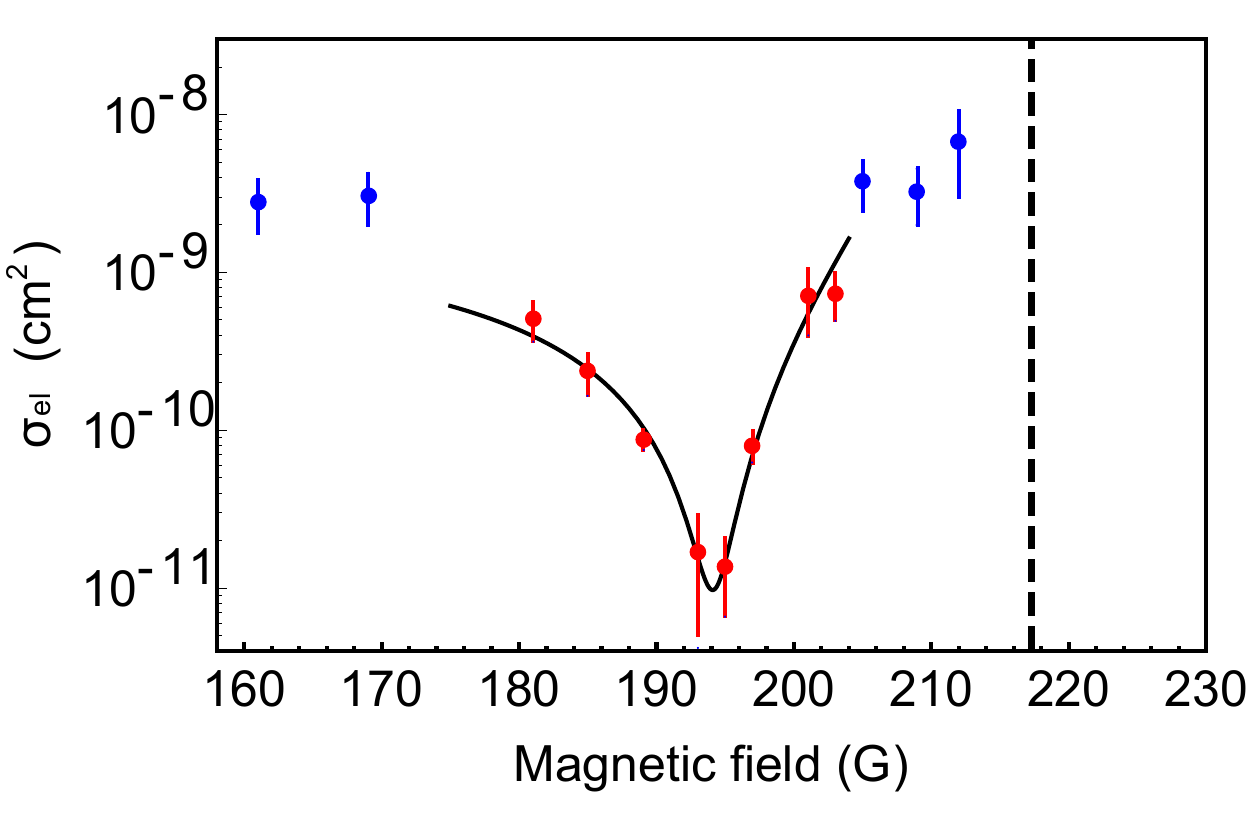}\caption{The elastic atom-molecule scattering cross sections extracted from the thermalization measurements. The red data points are fitted to the model $\sigma=4\pi|a|^2$ with the solid line. The dashed line is the position of the broad resonance at 217.3 G which is determined from the loss rate coefficients. The fitting parameters are $\Delta=-23.1(2)$ G, $a_{\mathrm{bg}}=-2902(182) a_0$, and $\gamma=2.7(5)$ G. Error bars shown in the plot are the standard errors.}
\label{Fig3.elastisection.pdf}
\end{figure}

A notable feature of this resonance is the unusually large background scattering length, which is similar to that of the $^{6}$Li atoms \cite{Abraham1997}. The reported background scattering lengths in fermionic atomic mixtures are less than 100 $a_0$ \cite{Wille2008, Tiecke2010,Ravensbergen2020,Green2020,Finelli2024,Florian2023}. The fitted $a_{\rm bg}$ in our system is larger than that in the fermionic atomic mixtures by more than one order of magnitude. Such a large negative background scattering length may arise from a virtual bound state of the three-body potential energy surface that lies very close to the threshold. It could also represent a local background scattering length associated with an even broader resonance. The large background scattering length combined with a width of 23.1 G indicates that this Feshbach resonance is highly open-channel dominated, with the strength parameter $A=a_{\rm bg}\Delta/a_0\approx 66000$ G being about one order of magnitude larger than that of the broad Feshbach resonance between $^{161}$Dy and $^{40}$K \cite{Ravensbergen2020}. We define a length parameter, $R=\hbar^2/(2m a_{\rm{bg}}\Delta\delta\mu)$ \cite{Petrov2004b}, where $\delta \mu$ is the relative magnetic moment. When the scattering length $|a| \gg R$, the universality condition is satisfied and the few-body collisional losses can be strongly suppressed by the Pauli principle. With a hypothetical value of $\delta \mu=1.5\ \mu_{B}$ which is the same as another Feshbach resonance at about 48.2 G \cite{Yang2022b}, we have $R=0.5 \ a_0$ relevant to this resonance. A more conservative estimate using $\delta \mu=0.1\ \mu_{B}$ yields $R=8\ a_0$, therefore the universality condition can be easily fulfilled.

The large background scattering length also implies strong interaction is accessible even when the magnetic field is far from the resonance. The strong interaction can drive the atom-molecule mixture into the hydrodynamic regime, where the collision rate is so large that multiple elastic collisions can lead to a hydrodynamic drag and an agreement between the flow velocities of the atoms and molecules \cite{Gensemer2001,Trenkwalder2011,Ravensbergen2020, Finelli2024}. The hydrodynamic behaviour can be examined by monitoring the motion of the center of mass. In the K-NaK mixture, the atomic and molecular motions of the center of mass can be described by a classical kinetic model
\begin{equation}
    \begin{aligned}
    \ddot y_\mathrm{K}&=-\omega_\mathrm{K}^2y_\mathrm{K}-\frac{4}{3}\frac{m_{\mathrm{NaK}}}{M}\frac{N_{\mathrm{NaK}}}{N}\Gamma_{\rm{coll}}(\Dot{y_\mathrm{K}}-\Dot{y_{\mathrm{NaK}}})\\
    \ddot y_{\mathrm{NaK}}&=-\omega_{\mathrm{NaK}}^2y_{\mathrm{NaK}}+\frac{4}{3}\frac{m_\mathrm{K}}{M}\frac{N_\mathrm{K}}{N}\Gamma_{\rm{coll}}(\Dot{y_\mathrm{K}}-\Dot{y_{\mathrm{NaK}}}),
    \end{aligned}
\end{equation}
where $M=m_{\mathrm{K}}+m_{\mathrm{NaK}}$ is the total mass and $N=N_{\mathrm{K}}+N_{\mathrm{NaK}}$ is the total particle number. In the above equation, the two motions are coupled through the elastic collisions. Under the circumstances of weak interactions where \(\Gamma_{\rm{coll}} < \omega_{\rm{K}}, \omega_{\rm{NaK}}\), the system is in the collisionless regime. The atomic and molecular clouds oscillate at their respective trap frequencies, with collisions contributing to a damping term. However in strong interaction case, i.e., \(\Gamma_{\rm{coll}} > \omega_{\rm{K}}, \omega_{\rm{NaK}}\), the atom-molecule mixture will step into the hydrodynamic regime. The clouds exert a mutual hydrodynamic drag, causing the motions of the center of mass to oscillate at a common frequency.

We investigate the hydrodynamic behaviour at 77.6 G, approximately \(7\Delta\) away from the resonance position. We excite the motions of the center of mass along the gravitational direction, where the trap frequency exceeds those of the other two directions. The excitation process is implemented by switching off the optical dipole trap, allowing the clouds to fall freely for 0.5 ms. We then switch on the dipole trap, enabling the atoms and molecules to slosh in the dipole trap for a variable interval. The particle clouds are then released from the trap, and the positions of the center of mass after a time of flight are measured. To mitigate systematic errors due to long-term drifts, measurements for atoms and molecules are performed alternately, with the results being an average of 6-8 measurements. The observed oscillations of center of mass are shown in Fig. 4. The results are fitted to a damped oscillator $y(t)=A e^{-t/\tau}\cos(\omega t+\phi)+B$. As a reference, we first measure the oscillation frequencies of pure molecules and atoms. The measured sloshing frequencies for the pure \(^{23}\)Na\(^{40}\)K and $^{40}$K samples are \(\omega_{\mathrm{NaK}} = 2\pi \times 205(2)\) Hz and \(\omega_{\mathrm{K}} = 2\pi \times 265(2)\) Hz, respectively. The ratio of the trap frequencies $\gamma = \omega_{\mathrm{NaK}} / \omega_{\mathrm{K}}=0.77$ is in good agreement with the theoretical calculation of 0.79 \cite{Vexiau2017}.

%However, in the atom-molecule mixture, the oscillation frequency of the molecules and the frequency of the atoms are the frequency is \textcolor{red}{$\omega^{'}_{\mathrm{K}}=2\pi\times254(3)$} Hz for atoms and \textcolor{red}{$\omega^{'}_{\mathrm{NaK}}=2\pi\times254(2)$} Hz. The two frequency agree with each other within experimental uncertainty. This clear demonstrates the mixture is in hydrodynamic regime, where the atoms and molecules oscillate with the same frequency due to the strong interactions. The above measurements also suggest that the oscillation frequencies of atoms is reduced by the strong interaction between atoms and molecules, even if the number of atoms is one orders of magnitude larger than the molecules. This is further confirmed by performing a separate experiment to directly compare the frequency of K atoms with and without the presence of NaK molecules, by measuring the alternately. The measured ratio $\omega^{'}_{\mathrm{K}}/\omega_{\mathrm{K}}=0.96$. This clearly demonstrates that the frequency of K atoms shifted due to the collisions with the NaK molecules. Note that our experiment is different from the experiments that the hydrodynamic effect is observed along the direction where the trap frequency is the weakest one of the three directions. Therefore, since the hydrodynamic effect is observed along the strongest direction, we may conclude that the motion of the two clouds are and thus cannot be separated. This indicates the strong interactions already.

\begin{figure}[ptb]
\centering
\includegraphics[width=8cm]{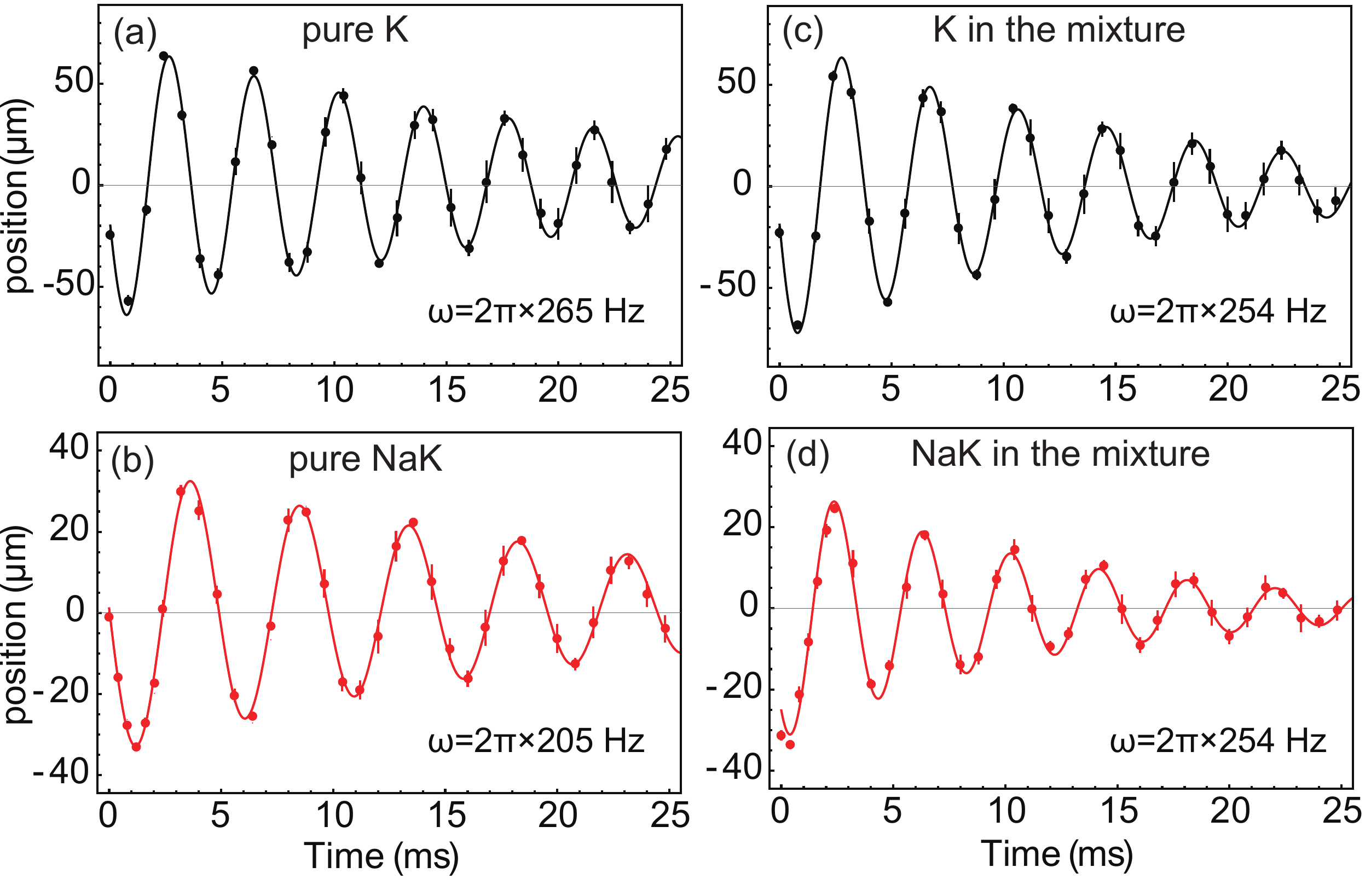}
\caption{The center of mass oscillations of atoms and molecules along the gravitational direction in the optical dipole trap. (a) and (b) are the results of pure atomic and molecular clouds, respectively. The two clouds oscillate with their own frequencies in the absence of atom-molecular interaction. (c) and (d) represent the oscillations of atomic and molecular clouds in the atom-molecule mixture. The two clouds oscillate with a common frequency due to the hydrodynamic drag effects. }
\label{Fig4.hydrodynamicdata.pdf}
\end{figure}

However, in the atom-molecule mixture, the oscillation frequencies of the motions of the center of mass are modified by collisions. The measured frequencies are $\widetilde{\omega}_{\mathrm{K}} = 2\pi \times 254(3)$ Hz and $\widetilde{\omega}_{\mathrm{NaK}} = 2\pi \times 254(2)$ Hz, which are consistent within the experimental uncertainty. It clearly demonstrates that the atom-molecule mixture is in the hydrodynamic regime, where the atoms and molecules oscillate at a common frequency due to the hydrodynamic drag. The observed oscillation frequency is smaller than the theoretical calculations \cite{supp} by a few percent, which may be due to the long-term drift of the trap frequency. Given the typical molecular lifetime of about 40 ms in the mixture, we only measure the oscillation along the gravitational direction, where more periods can be observed than the other two directions. Since the gravitational direction has the tightest confinement, if hydrodynamic drag effects are observed along this direction,
%The measurements further suggest that the oscillation frequency of the atoms is influenced by the strong atom-molecule interaction, despite the atomic density being approximately one order of magnitude higher than that of the molecules.
%The measured ratio $\widetilde{\omega}_{\mathrm{K}} / \omega_{\mathrm{K}} = 0.97$ means that the sloshing frequency of $^{40}$K atoms has shifted by due to collisions with the $^{23}$Na$^{40}$K  molecules.
%Note that our experiment differs from those works where the hydrodynamic effect is observed along the direction where the trap frequency is the weakest of the three directions.
%Note that in our experiment, hydrodynamic drag effects are observed along the direction with the highest trap frequency. This is distinct from those experiments where hydrodynamic behaviour is observed along the axial direction with the lowest trap frequency \cite{Gensemer2001,Finelli2024}.
%{\color{red}Since the typical lifetime of the molecules in the mixture is about 40 ms, we only measure the oscillation along the gravitational direction, where more periods can be observed than the other two directions. Because the gravitational direction is the direction with the tightest confinement, if hydrodynamic drag effects are observed along the direction,
we may conclude that the atom-molecule mixture exhibits hydrodynamic behavior in all three spatial directions. Consequently, the observed hydrodynamic drag effects may be regarded as a hallmark of the strongly interacting regime, analogous to the hydrodynamic expansion observed near the Feshbach resonances between $^{6}$Li and $^{40}$K \cite{Trenkwalder2011}, and $^{161}$Dy and $^{40}$K \cite{Ravensbergen2020}.

In conclusion,  we have observed a broad Feshbach resonance with a large background scattering length between $^{23}$Na$^{40}$K and $^{40}$K in their absolute ground states.  This resonance may be used to study mass-imbalanced BCS-type fermionic superfluid \cite{Iskin2006,Hanai2014,Pini2021}. The large universal range is particularly useful to study the BEC-BCS crossover. Other exotic phases such as interior gap \cite{Liu2003} and FFLO phases \cite{Fulde1964,Larkin1965,WangJ2021} might also be explored using this broad resonance.
%The large background scattering length enables the mixture to enter the hydrodynamic regime even at magnetic fields far from resonance. The large universal range renders this resonance a prospective candidate for exploring mass-imbalanced strongly interacting fermionic gases, as the Pauli exclusion principle can help suppress few-body collisional instabilities.
However, the short lifetime of molecules near the resonance is a critical issue, because the study of BEC-BCS crossover requires a long lifetime near the resonance. The molecular loss near the resonance is mainly caused by the photoexcitation by the 1064 nm trapping laser \cite{Nichols2022,Cao2024}.  The trapping laser can couple the atom-molecule scattering state to the triatomic molecules in the electronically excited state, whose spectrum is quasi-continuous due to the high density of states. This problem can be mitigated by increasing the wavelength of the trapping laser. In Ref. \cite{Cao2024}, we have observed that the lifetime of the molecule in the mixture near the Feshbach resonances can be enhanced by using 1558/1583 nm optical dipole trap and tuning the laser frequency off-resonant from photoassociation transitions. The \emph{ab initio} quantum chemistry calculations reveal that the minimum of the electronically excited state on the three-body potential energy surface lies about 3000 cm$^{-1}$ above the atom-molecule collision threshold \cite{Cao2024}. Therefore, using trapping lasers with wavelengths larger than 3 $\mu$m may fully suppress the photoexcitations in atom-molecule collisions. In this scenario, three-body recombination is expected to be the dominant loss mechanism. As both the atoms and the molecules are fermionic particles, such losses can in principle be suppressed in the vicinity of the broad Feshbach resonance owing to the Pauli exclusion principle. The few-body collisional stability needs to be studied to unlock the potential of the broad resonance.
%{\color{red}Using a long wavelength of several microns as the trapping laser may also be helpful to suppress the losses in molecule-molecule collisions, as has been proposed in Ref. \cite{Christianen2019}}

%Through the investigation of the inelastic and elastic collision process in detail, we characterize this broad Feshbach resonance directly. The experiment results show that this scattering resonance in $^{23}$Na$^{40}$K+$^{40}$K collisions can widely vary the interaction while the large background scattering length and large resonance width are favorable for the study of strongly interacting fermionic systems. Especially, the large background scattering cross-section gives rise to the hydrodynamic behavior of the $^{23}$Na$^{40}$K-$^{40}$K sample. With this broad resonance, the ultracold atom-molecule system holds promise to achieving the exotic interaction regimes, in particular concerning the novel superfluid states in the atom-diatomic molecule system and the Bose-Einstein condensation of triatomic molecules.

\begin{acknowledgments}
This work was supported by the National Natural Science Foundation of China (under Grant No. 12241409, 12325407, 12422409, 12488301 and 12274393), the Chinese Academy of Sciences, the Shanghai Municipal Science and Technology Major Project (Grant No. 2019SHZDZX01), the Quantum Science and Technology-National Science and Technology Major Project (Grant No. 2021ZD0302101).
\end{acknowledgments}
Data availability: The data that support the findings of this article are openly available \cite{dataset}.

\section{Supplementary materials}

\section*{Search for broad Feshbach resonances}

\noindent

After preparing an ultracold mixture of $^{23}$Na$^{40}$K molecules and $^{40}$K atoms at 77.6 G, we search for atom-molecule Feshbach resonances as follows. We ramp the magnetic field to a target value within 15 ms and hold the magnetic field at the target value for 20 ms. The hold time of the mixture is controlled by removing the $^{40}$K atoms with a resonant light pulse. The magnetic field is then ramped back to 77.6 G and the remaining number of molecules is recorded. The number of molecules for a hold time of 5 ms, normalized to the molecule number for a hold time of 0 ms, is recorded. In our previous works, we have searched for atom-molecule resonances in the range between 16 G and 120 G at a step of about 0.5 G. In this work, we start from 120 G and increase the target magnetic field with a step of about 1-2 G, until a broad Feshbach resonance at 217 G is observed.

To demonstrate the loss features are caused by atom-molecule resonances, we also measure the loss of pure molecules in the vicinity of the broad Feshbach resonance. For this measurement, we first remove the $^{40}$K atoms and then ramp the magnetic field to a target value within 50 ms and hold the magnetic field at the target value for 10 ms or 20 ms. The number of molecules for a hold time of 20 ms, normalized to the molecule number for a hold time of 10 ms, is measured. No loss features are observed for pure molecules. This confirms the loss features are caused by the atom-molecule Feshbach resonances. A comparison of atom-molecule mixture and pure molecules is shown in Fig. S1.

\begin{figure}[ptb]
\centering
\renewcommand\thefigure{S1}
\includegraphics[width=8cm]{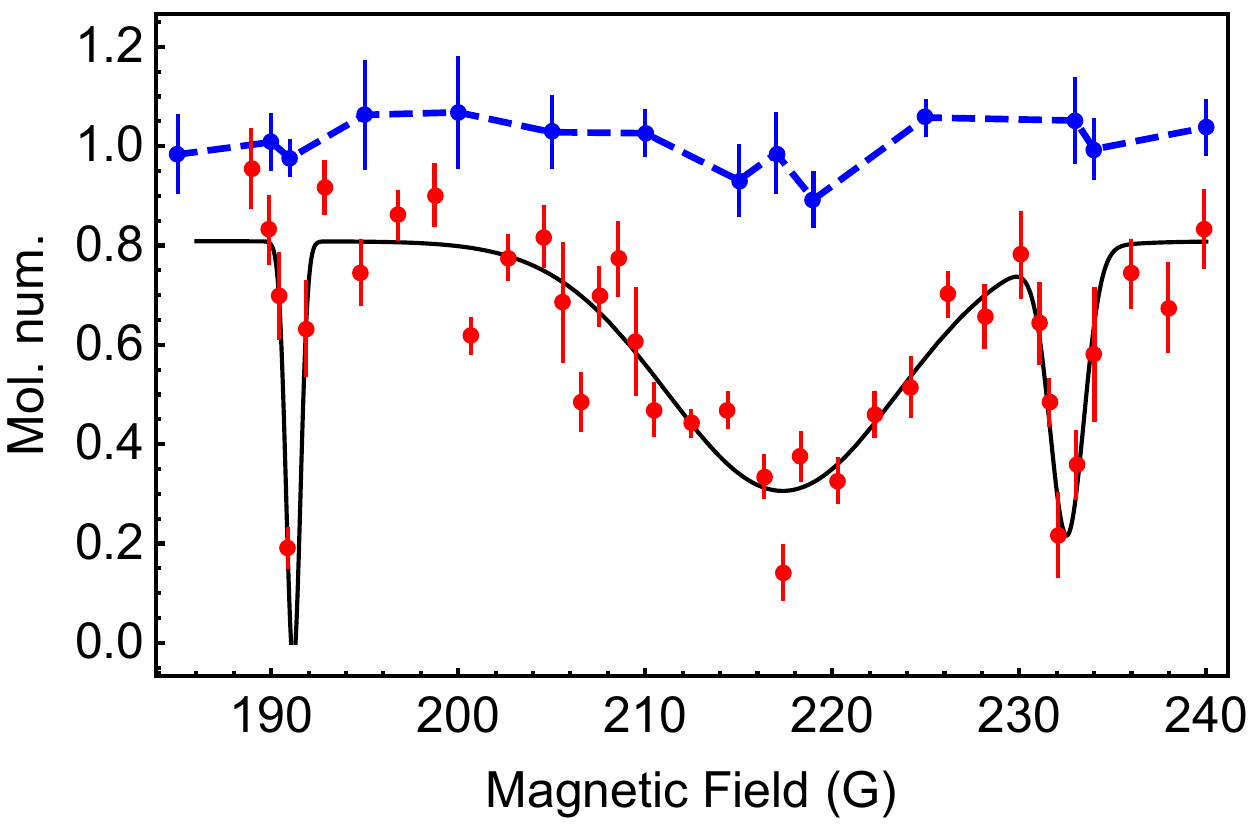}
\caption{Comparison of loss spectrum of pure molecules (blue points) and molecules in the atom-molecule mixture (red points). Error bars represent the standard error of the mean.}%
\label{fig4}%
\end{figure}

\section*{The elastic scattering cross sections}

%For a Maxwell-Boltzmann distribution, the column density distributions obtained from the absorption image can be fit to two-dimensional Gaussian functions $\propto e^{-x^2/(2\sigma_x^2)-y^2/(2\sigma_y^2)}$, where the cloud sizes along the $x$ and $y$ directions can be described by  $\sigma_{x,y}^2(t)=[k_{\rm{B}}T(t)/m_{\rm{NaK}}](\tau_{\rm{tof}}^2+1/\omega_{x,y}^2)$, with $\tau_{\rm{tof}}$ being the time of flight.

In the analysis of the elastic scattering cross sections in the main text, we have neglected the effects of the gravitational sag between $^{40}$K atoms and $^{23}$Na$^{40}$K molecules, which is about 2.4 $\mu$m. The calculated RMS in-trap sizes of the atomic cloud and the molecular cloud along the gravitational direction are about 6.7 $\mu$m and 6.9 $\mu$m, respectively. Therefore, we estimate that neglecting the gravitational sag will cause a systematic error of about 4\%. This is smaller than the uncertainties of the elastic scattering cross sections, which are typically about 30\%-40\%.

In the analysis in the main text, we have attributed the resonance to a resonance in \emph{s}-wave scattering. This is justified from the measurement of the minimum of the elastic scattering cross sections. If the resonance is associated to a high partial wave, e.g., \emph{p}-wave, the total elastic scattering cross section should be equal to the sum of the \emph{s}-wave and \emph{p}-wave scattering cross sections. The contribution from the \emph{p}-wave scattering will manifest itself at resonance, while the \emph{s}-wave scattering will contribute to a background scattering cross section. In this case, at the magnetic field where the \emph{p}-wave scattering cross section is minimum, the total elastic scattering cross section should be equal to the \emph{s}-wave background scattering cross section. This means the minimum total scattering cross section at about 194 G should be equal the background scattering cross section far from the resonance, because the contribution of \emph{p}-wave scattering is be negligible far from resonance at ultralow temperatures. However, in the experiment, the minimum scattering cross section at 194 G is much smaller than the background scattering cross sections far from the resonance. Therefore, we may conclude that the resonance is an \emph{s}-wave resonance.

There are two resonances close to the minimum of the elastic scattering cross section at 194 G. One is the narrow resonance at 191 G and the other one is the broad resonance at 217 G. We attribute the minimum at 194 G to the resonance at 217 G due to the following reasons. First, the resonance at 191 G is very narrow. For a narrow resonance, the range of the magnetic field where the scattering cross sections change from minimum to the background value is also narrow. However, in our experiment, the range where the scattering cross sections change from minimum to the background value is on the order of 10 G. This is much larger than the width of the resonance at 191 G. Therefore, the minimum should be associated to a broad resonance but not a narrow resonance. Second, when we fit the data points near the minimum of the scattering cross sections in Fig. 3 in the main text, the resonance position is fixed to be 217.3 G. If we change the resonance position to 191 G, the curve significantly deviates from the data and the fitting result has a large uncertainty. Due to these reasons, we attribute the minimum at 194 G to the broad resonance at 217 G.

\begin{figure}[ptb]
\centering
\renewcommand\thefigure{S2}
\includegraphics[width=8cm]{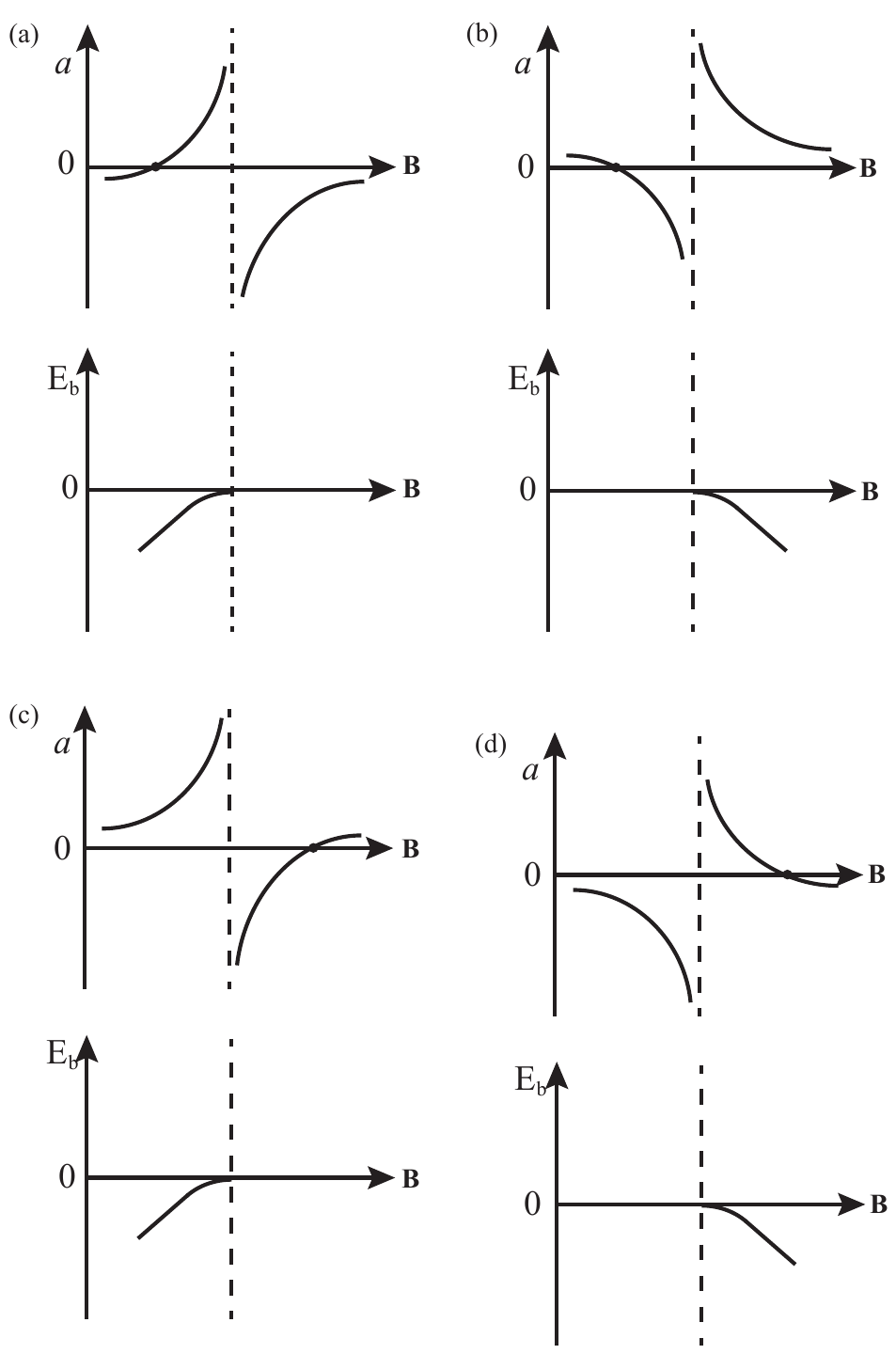}
\caption{(a) $a_{\rm{bg}}$ is negative, the magnetic field where the scattering cross section is minimum is smaller than the resonance position, the bound state approaches the resonance from the low field side; (b) $a_{\rm{bg}}$ is positive, the magnetic field where the scattering cross section is minimum is smaller than the resonance position, the bound state approaches the resonance from the high field side; (c) $a_{\rm{bg}}$ is positive, the magnetic field where the scattering cross section is minimum is larger than the resonance position, the bound state approaches the resonance from the low field side; (d) $a_{\rm{bg}}$ is negative, the magnetic field where the scattering cross section is minimum is larger than the resonance position, the bound state approaches the resonance from the high field side.}%
\label{fig4}%
\end{figure}

The background scattering length is only associated with the open channel and is independent from the closed channel. However, in the vicinity of a Feshbach resonance, the scattering length will change from the background value to infinity or minus infinity as a function of the magnetic field. During this process, the scattering length will cross zero at a certain magnetic field. Therefore, there is a close relation between the sign of the background scattering length, the magnetic field where the scattering cross section is minimum, and from which side the bound state approaches the resonance. As shown in Fig. S2, there are in total four possibilities: (a) $a_{\rm{bg}}$ is negative, the magnetic field where the scattering cross section is minimum is smaller than the resonance position, and the bound state approaches the resonance from the low field side; (b) $a_{\rm{bg}}$ is positive, the magnetic field where the scattering cross section is minimum is smaller than the resonance position, and the bound state approaches the resonance from the high field side; (c) $a_{\rm{bg}}$ is positive, the magnetic field where the scattering cross section is minimum is larger than the resonance position, and the bound state approaches the resonance from the low field side; (d) $a_{\rm{bg}}$ is negative, the magnetic field where the scattering cross section is minimum is larger than the resonance position, and the bound state approaches the resonance from the high field side.

In our experiment, we have observed that the magnetic field where the scattering cross section is minimum is smaller than the resonance position. Therefore, only cases (a) and (b) are possible. Thus, to determine the sign of the background scattering length, we only need to know whether the triatomic molecule bound state approaches the resonance from the low field side or the high field side. For the collision channel where the atom is in the $|9/2, -9/2\rangle$ state and the molecule is in the lowest hyperfine state $|0,0,3/2,-4\rangle$, if the triatomic molecule approaches the scattering state from the high field side, the slope of Zeeman shift of triatomic state must be negative and the absolute value of the slope should be larger than the slope of the scattering state, which may be described by
\begin{align}
|g_s m_s \mu_B+(g_{I_{\rm{Na}}} m_{I_{\rm{Na}}}+g_{I_{\rm{K1}}} m_{I_{\rm{K1}}}+g_{I_{\rm{K2}}} m_{I_{\rm{K2}}})\mu_N| \notag\\
=|-\frac{1}{2}g_s \mu_B+(\frac{3}{2}g_{I_{\rm{Na}}} -4 g_{I_{\rm{K1}}} -4 g_{I_{\rm{K2}}} )\mu_N|  \notag
\end{align}
where $g_s, g_{I_{\rm{K1}}}g_{I_{\rm{K2}}}>0$, and $g_{I_{\rm{Na}}}<0$. The electronic spin of the triatomic molecule is $S=1/2$, and the nuclear spin of triatomic molecule includes the contributions from one $^{23}$Na atom $I_{\rm{Na}}=3/2$ , and two $^{40}$K atoms $I_{\rm{K1}}=I_{\rm{K2}}=4$. If the slope of the Zeeman shift of triatomic molecule is negative, the maximum absolute value of the slope is also $|-\frac{1}{2}g_s \mu_B+(\frac{3}{2}g_{I_{\rm{Na}}} -4 g_{I_{\rm{K1}}} -4 g_{I_{\rm{K2}}} )\mu_N|$. Therefore, the slope of triatomic molecule state cannot be larger than that of the slope of the scattering state. According to these arguments, we may conclude that the triatomic molecule state should approach the scattering state from the low field side and thus the background scattering length $a_{\rm{bg}}$ is negative.

\section*{Theoretical calculation of the oscillation frequency in the hydrodynamic regime}

The center-of-mass oscillation of the $^{40}$K atoms and $^{23}$Na$^{40}$K molecules is described by Eq. (2) in the main text. The oscillation frequency can be calculated by solving the equation $Det|A|=0$, where
\begin{equation}
A=\left(
  \begin{array}{cc}
    \omega^2-\omega_\mathrm{K}^2-i\alpha\Gamma_{\rm{coll}} \omega & i\alpha\Gamma_{\rm{coll}}\omega  \notag\\
    i\beta \Gamma_{\rm{coll}} \omega & \omega^2-\omega_\mathrm{NaK}^2-i\beta\Gamma_{\rm{coll}} \omega \notag\\
  \end{array}
\right)
\end{equation}
with $\alpha=\frac{4}{3}\frac{m_{\mathrm{NaK}}}{M}\frac{N_{\mathrm{NaK}}}{N}$ and $\beta=\frac{4}{3}\frac{m_{\mathrm{K}}}{M}\frac{N_{\mathrm{K}}}{N}$. The real part of the solution gives the frequency of the oscillation and the imaginal part gives the damping rate. The bare frequencies are $\omega_{\rm{K}}=2\pi\times 265$ Hz and $\omega_{\rm{NaK}}=2\pi\times205$ Hz.
%, with $M=m_{\mathrm{NaK}}+m_{\mathrm{K}}$ and $N=N_{\mathrm{NaK}}+m_{\mathrm{K}}$, $\Gamma_{\rm{coll}}=n\sigma v$
The collision rate is calculated by $\Gamma_{\rm{coll}}=n_{\rm{ov}}\sigma_{\rm{el}} v_{\rm{rel}}$. Here, $n_{\rm{ov}}=(N_{\rm{NaK}}+N_{\rm{K}})[(\frac{2\pi k_{\rm{B}} T_{\rm{K}}}{m_{\rm{K}} \bar\omega_{\rm{K}}^2})(1+\frac{m_{\rm{K}}T_{\rm{NaK}}}{m_{\rm{NaK}} T_{\rm{K}} \gamma_t^2} )]^{-\frac{3}{2}}$ represents the overlap density with $\gamma_t=0.77$ being the ratio between the trap frequencies for $^{23}$Na$^{40}$K molecules and $^{40}$K atoms, $\sigma_{\rm{el}}=4\pi|a_{\rm{bg}}|^2$ is the background elastic scattering cross sections, and $v_{\rm{rel}}=\sqrt{(8 k_{\rm{B}}/\pi)(T_{\rm{K}}/m_{\rm{K}}+T_{\rm{NaK}}/m_{\rm{NaK}})}$ is the relative mean velocity. Given the parameters $N_{\rm{K}}=2.8\times 10^5$, $N_{\rm{NaK}}=1.4\times 10^4$, $T_{\rm{K}}=T_{\rm{NaK}}=300$ nK, $a_{\rm{bg}}=-2902$ $a_0$, we obtain $\Gamma_{\rm{coll}}\approx2\pi\times 1083$ Hz. In the hydrodynamics regime, only the phase-locked oscillation exists and we calculate the frequency of the phase-locked oscillation $\omega\approx 2\pi\times 261 $ Hz. The theoretical calculated frequency is slightly higher than the measured frequency $\widetilde{\omega}_{\mathrm{K}} = 2\pi \times 254(3)$ Hz and $\widetilde{\omega}_{\mathrm{NaK}} = 2\pi \times 254(2)$ Hz by a few percent. We attribute this discrepency to the long term drift of the trap frequency.

\end{document}